\documentclass[12pt]{article}
\usepackage{amsmath,amssymb}
\usepackage{graphicx}
\def\R{\hbox{{\rm I}\kern-0.2em{\rm R}\kern0.2em}}
\def\R{\hbox{{\rm I}\kern-0.2em{\rm R}\kern0.2em}}
\def\D{\hbox{{\rm I}\kern-0.2em{\rm D}\kern0.2em}}

\def\e{{\rm e}}

\def\be{\begin{equation}}
\def\ee{\end{equation}}

\def\({\left(}
\def\){\right)}
\def\[{\left[}
\def\]{\right]}
\def\bc{\begin{center}}
\def\ec{\end{center}}

\begin{document}

{\large \bf A Proposal for Determining the Energy Content of
Gravitational Waves by Using Approximate Symmetries of Differential
Equations} \footnote {Dedicated to the memory of John Archibald
Wheeler whose ``poor man" approaches yielded such rich insights.}

\textit{IBRAR HUSSAIN}$^\dag$\footnote {Correspondence should be
addressed to ihussain@camp.edu.pk}, \textit{F. M. MAHOMED}$^\ddag$
and \textit{ASGHAR QADIR}$^{\dag}$

$^{\dag}$Centre for Advanced Mathematics and Physics (CAMP)\\
National University of Sciences and Technology\\
Campus of the College of Electrical and Mechanical Engineering\\
Peshawar Road, Rawalpindi, Pakistan

E-mail: ihussain@camp.edu.pk, aqadirmath@yahoo.com

$^{\ddag}$Centre for Differential Equations, Continuum Mechanics and Applications\\
School of Computational and Applied Mathematics (DECMA)\\
University of the Witwatersrand\\
Wits 2050, South Africa

E-mail: Fazal.Mahomed@wits.ac.za

{\bf Abstract}. Since gravitational wave spacetimes are time-varying
vacuum solutions of Einstein's field equations, there is no
unambiguous means to define their energy content. However, Weber and
Wheeler had demonstrated that they do impart energy to test
particles. There have been various proposals to define the energy
content but they have not met with great success. Here we propose a
definition using ``slightly broken" Noether symmetries. We check
whether this definition is physically acceptable. The procedure
adopted is to appeal to ``approximate symmetries" as defined in Lie
analysis and use them in the limit of the exact symmetry holding. A
problem is noted with the use of the proposal for plane-fronted
gravitational waves. To attain a better understanding of the
implications of this proposal we also use an artificially
constructed time-varying non-vacuum metric and evaluate its Weyl and
stress-energy tensors so as to obtain the gravitational and matter
components separately and compare them with the energy content
obtained by our proposal. The procedure is also used for cylindrical
gravitational wave solutions. The usefulness of the definition is
demonstrated by the fact that it leads to a result on whether
gravitational waves suffer self-damping.


\section{Introduction}

Gravitational wave spacetimes are non-static vacuum solutions of the
Einstein Field Equations (EFEs), i.e. they have no timelike Killing
Vectors (KVs) \cite{MTW}. This creates a problem with the definition
of energy for gravitational waves in General Relativity (GR), as
energy conservation is guaranteed for spacetimes that admit timelike
KVs. Since for these waves the stress-energy tensor is zero, there
was a debate whether they {\it really} exist \cite{WW, EK}. To
demonstrate their reality, Weber and Wheeler obtained (first and
second order) approximate formulae for the momentum imparted to test
particles in the path of cylindrical gravitational waves \cite{WW,
WW1}.  Later Ehlers and Kundt did the same for plane gravitational
waves \cite{EK}. Using the pseudo-Newtonian formalism Qadir and
Sharif \cite{QS} obtained a general closed form expression for the
momentum imparted to test particles in an arbitrary spacetime, which
gave the Weber-Wheeler approximation for cylindrical waves.

To define the energy content of gravitational waves, different
people have attempted different \textit{approximate symmetry}
approaches. One such attempt was to assume that conservation of
energy holds \textit{asymptotically} and examine whether it would
work for gravitational waves assuming a positive definite energy
\cite{KA}. An altogether different idea was adopted to provide a
measure of the extent of break-down of symmetry by the integral of
the square of the symmetrized derivative of a vector field divided
by its mean square norm \cite{MR,IR}. This led to what was called
an \textit{almost symmetric space} and the corresponding vector
field an \textit{almost Killing vector} \cite{YJ}. This measure of
``non-symmetry'' in a given direction was applied to the Taub
cosmological solution \cite{TA} and to study gravitational
radiation. It provides a choice of gauge that makes calculations
simpler and was used for this purpose \cite{BCP}. Essentially based
on the almost symmetry, the concept of an ``approximate symmetry
group'' was presented \cite{SB}. However none of them seem
unequivocally successful. The approach of a ``slightly broken
symmetry" seems promising but merely providing simplicity of
calculations is not physically convincing. Other approaches need to
be tried to find one that is significantly better than the others,
in that it is consistent with other physical concepts and leads to
new physical insights. It was speculated that the use of approximate
Lie symmetry methods for differential equations (DEs) \cite{BGI,Ib}
may give a resolution to the problem of energy in non-static
spacetimes \cite{IMQ, IMQ1}. In this paper we apply these methods to
propose a resolution of the problem of the energy content of
gravitational wave spacetimes.

By virtue of Noether's theorem \cite{Nth} for every infinitesimal
generator of symmetry of a Lagrangian (called a {\it Noether
symmetry}), there is a conserved quantity. For time translational
invariance it is energy that is conserved. It is for this reason
that it was hoped that the symmetry approach could prove fruitful
for defining the energy content of gravitational waves. Reducing
from the maximally symmetric Minkowski spacetime (10 KVs) to the
Schwarzschild \cite{KMQ} and Reissner-Nordstr\"{o}m (RN) \cite{IMQ}
spacetimes linear and spin angular momentum conservation are lost.
Using approximate Lie symmetry methods for DEs these conservation
laws were recovered as trivial first-order and second-order
approximate conservation laws respectively. Reducing from the
Schwarzschild spacetime to the Kerr (or charged-Kerr) spacetime we
lose angular momentum conservation. Going directly from the
Minkowski spacetime to the Kerr (or charged-Kerr) spacetime we lose
linear and spin angular momentum conservation as well. These lost
conservation laws were recovered as ``trivial'' first-order and
second-order approximate conservation laws for this spacetime
\cite{IMQ1} (in that there is no exact symmetry part mixed in with
the approximate symmetry infinitesimal generator).

The trivial first order approximate symmetries did not provide any
new insights. However, for the second-order approximate symmetries
of the geodesic equations for the RN \cite{IMQ} and charged-Kerr
\cite{IMQ1} spacetimes the time translational approximate symmetry
generator had to pick up a rescaling factor to provide a necessary
cancelation. This corresponds to rescaling the energy of test
particles and hence could possibly lead to a definition of energy in
the spacetime. Clearly, the energy {\it content} would then be
defined as the scaling factor.

To check the proposal for defining energy in gravitational wave
spacetimes by using {\it slightly broken symmetry} as defined in Lie
analysis \cite{BGI}, plane-fronted parallel-rays (pp) gravitational
waves \cite{MM} are first investigated. For the approximate
symmetries of pp-waves first the $t$-dependent part of this
spacetime is removed to make it static and taken as the unperturbed
spacetime. Then the exact pp-wave is taken as a perturbation on this
static spacetime by considering the arbitrary amplitude of the wave
as a small parameter, $\epsilon$. Since $\epsilon^2$ does not appear
in the geodesic equations for perturbed pp-waves, there is a problem
in applying the definition of second-order approximate symmetries of
ordinary differential equations (ODEs), which gives the scaling
factor mentioned earlier, to them. It can also be seen from the
geometry of pp-waves in which the wave fronts are like moving
parallel planes and the curvature of the spacetime is absolutely
zero before the pp-wave pulse arrives and after it has passed
\cite{MTW}. There is no region where there is a {\it slight} shift
from the flat geometry as required for obtaining an approximate
symmetry. Thus the proposal for determining the energy content of
pp-waves cannot be checked. The conformally invariant Weyl tensor
\cite{HkEl} represents a pure gravitational field. In some sense it
tells us about the gravitational energy of the spacetime, but it
does not give a direct measure of the gravitational energy. The
stress-energy tensor gives the matter content of the spacetime
\cite{MTW}. For this perturbed spacetime the stress-energy tensor
is zero while the Weyl tensor is nonzero. To obtain a better
understanding of the energy rescaling in plane gravitational waves,
the artificially constructed example of a plane symmetric
``wave-like" spacetime \cite{IA}, which represents a gravitational
wave interacting with matter is investigated. Here we {\it do} obtain
a scaling factor which gives the rescaling of energy in the spacetime
field.

For the plane wave-like spacetime \cite{IA}, along with trivial
approximate symmetries, a non-trivial first-order approximate
symmetry was found. The first-order approximate (stable) first
integral corresponding to the non-trivial first-order approximate
symmetry of this wave-like spacetime is calculated here. The
first-order non-trivial approximate time-like Noether symmetry is
used with the momentum vector to obtain a conserved quantity which
gives the energy non-conservation due to time variation. To check
this quantity first-order approximate Noether symmetries of pp-wave
spacetime are investigated. Only the exact symmetries are recovered
as trivial first-order approximate Noether symmetries, which gives
the exact conservation laws as trivial first-order approximate
conservation laws.

Next cylindrically symmetric exact gravitational waves \cite{ER},
which are physically easier to understand, are considered. To study
their approximate symmetries the $t$-dependent part is first removed
to define a static spacetime and the exact wave is dealt with as a
perturbation of this static spacetime, taking the strength of the
wave as a small parameter, $\epsilon$. A scaling factor is obtained
for this spacetime which gives the rescaling of energy. Noether
symmetries for the cylindrically symmetric case are also considered
to look at the conserved quantities. First a cylindrically symmetric
wave-like spacetime is investigated which has a non-trivial
first-order approximate Noether symmetry that gives the conserved
quantity like the plane symmetric case. There is no non-trivial
approximate symmetry for the perturbed cylindrical wave spacetime.
The approximate Weyl and stress-energy tensors (up to first-order in
$\epsilon$) for the cylindrical wave spacetimes are nonzero.
Electromagnetic waves in their interaction with matter get damped.
This is known as Landau damping \cite{Lan}. Since GR is a non-linear
theory, gravitational waves have self-interactions. The question
arises whether we should expect Landau self-damping. With our
proposal, a self-damping is seen for cylindrical waves.

The plan of the paper is as follows. The next section briefly
reviews the mathematical formalism to be used. In section 3,
second-order approximate symmetries of the geodesic equations for
the plane wave spacetimes are discussed and graphs of the
scaling factor for them are given. In the same section the Weyl
and stress-energy tensors for plane wave spacetimes are also
discussed. Noether symmetries of pp-waves are studied
and a review of the approximate Noether symmetries for the plane
symmetric wave-like spacetime is given, in section 4. In the next
section, second-order approximate symmetries of the geodesic
equations for the cylindrically symmetric case are provided and the
graphs of the scaling factor for them given. The Weyl and
stress-energy tensors for the cylindrical waves are also discussed
in the same section. In section 6 approximate Noether symmetries for
the cylindrical wave spacetimes are investigated. Finally a summary
and discussion are presented in section 7.

\section{Review of mathematical formalism used}

We first review the definition of the second-order approximate
symmetries of a system of ODEs under point symmetries. A vector
field
\begin{equation}
\mathbf{X}=\mathbf{X}_{0}+\epsilon\mathbf{X}_{1}+\epsilon^{2}\mathbf{X}_{2}+O(\epsilon^3),
\end{equation}
is called a second-order approximate symmetry of the system of
perturbed ODEs
\begin{equation}
\mathbf{E}=\mathbf{E}_{0}+\epsilon\mathbf{E}_{1}+\epsilon^{2}\mathbf{E}_{2}+O(\epsilon^3),
\end{equation}
if the following condition (\cite{IMQ} and references given there
in) holds
\begin{equation}
(\mathbf{X}_{0}+\epsilon\mathbf{X}_{1}+\epsilon^{2}\mathbf{X}_{2})\left(\mathbf{E}_{0}
+\epsilon\mathbf{E}_{1}+\epsilon^{2}\mathbf{E}_{2})\right\vert_{\mathbf{E}_
{0}+\epsilon\mathbf{E}_{1}+\epsilon^{2}\mathbf{E}_{2}=O(\epsilon^{3})}=O(\epsilon^{3}),
\end{equation}
where $\mathbf{X}_{0}$ is the exact symmetry generator of the system
of ODEs ${\bf E_0}$, i.e.
\begin{equation}
(\mathbf{X}_{0})\left(\mathbf{E}_{0})\right\vert_{\mathbf{E}_{0}=0}=0,
\end{equation}
$\mathbf{X}_{1}$, $\mathbf{X}_{2}$ are the first-order and
second-order approximate parts of the approximate symmetry generator
respectively, $\mathbf{E}_{1}$ is the first-order perturbed part and
$\mathbf{E}_{2}$ is the second order perturbed part of the system of
ODEs respectively. It should be noted that the scaling factor comes
from the applications of the perturbed system of DEs in subscript of
(3), as required.

Noether symmetries are those infinitesimal symmetry generators that
leave a Lagrangian $L(s, x^{\mu}, \dot{x}^{\mu})$ invariant. They
form a Lie algebra that contains the isometries for the Lagrangian
that minimizes arc length, with at least one extra symmetry,
${\partial}/{\partial s}$, \cite{AQ}. It is defined as a vector
field \cite{Ib, WF, KTT}
\begin{equation}
\mathbf {X}\mathbf{=}\xi (s,{x}^{\mu})\frac{\partial }{\partial
s}+\eta^{\nu}(s,{x}^{\mu})\frac{\partial}{\partial {x}^{\nu}},
\end{equation}
where $\mu, \nu=0,1,2,3$, such that
\begin{equation}
\mathbf{X}^{[1]}L+(D_{s}\xi)L=D_{s}A,
\end{equation}
where $A(s,x^{\mu})$, is a gauge function. The total derivative
operator $D_s$ and the first prolongation $\mathbf{X}^{[1]}$ of the
vector field ${\bf X}$ given by (5) are
\begin{equation}
D_{s}=\frac{\partial }{\partial s}+\dot{x}^{\mu}\frac{\partial}
{\partial x^{\mu}},
\end{equation}
and
\begin{equation}
\mathbf{X}^{[1]}=\mathbf{X}+(\eta^{\nu}_{,s}+\eta^{\nu}_{,{\mu}}\dot{x}^{\mu}
-\xi_{,s}\dot{x}^{\nu}-\xi_{,{\mu}}\dot{x}^{\mu} \dot{x}^{\nu})
\frac{\partial}{\partial \dot{x}^{\nu}}.
\end{equation}
For more general considerations see \cite{Ib}. The significance of
Noether symmetries is clear from the following theorem \cite{Nth},
proved in \cite{Ovs}.

{\bf Theorem 1}. If {\bf X} is a Noether point symmetry
corresponding to a Lagrangian $L(s,x^{\mu},\dot{x}^{\mu})$ of a
second-order ODE $\ddot{x}^{\mu}=g(s, x, \dot{x}^{\mu})$, then
\begin{equation}
I={\xi}L+(\eta^{\mu}-\dot{x}^{\mu}{\xi})\frac{\partial L}{\partial
\dot{x}^{\mu}}-A, \label{6}
\end{equation}
is a first integral of the ODE associated with {\bf X}.

First-order approximate symmetries of a Lagrangian (or first-order
approximate Noether symmetries) \cite{KTT, TK} are defined as
follows. For a first-order perturbed system of ODEs
\begin{equation}
\mathbf{E}=\mathbf{E}_{0}+\epsilon\mathbf{E}_{1}=O({\epsilon}^{2}),
\end{equation}
corresponding to a first-order Lagrangian, which is perturbed up to
first-order in $\epsilon$,
\begin{equation}
L(s,x^{\mu},\dot{x}^{\mu},\epsilon)=L_{0}(s,x^{\mu},\dot{x}^{\mu})+
{\epsilon}L_{1}(s,x^{\mu},\dot{x}^{\mu})+O({\epsilon}^{2}),
\end{equation}
the functional ${\int}_{V}{L}ds$ is invariant under the one-parameter
group of transformations with approximate Lie symmetry generator
\begin{equation}
\mathbf{X}=\mathbf{X}_{0}+\epsilon\mathbf{X}_{1} +
O({\epsilon}^{2}),
\end{equation}
up to gauge
\begin{equation}
A={A}_{0}+\epsilon{A}_{1},
\end{equation}
where
\begin{equation}
\mathbf{X}_{j}=\xi_{j}\frac{\partial}{\partial
s}+\eta^{\mu}_{j}\frac{\partial }{\partial {x}^{\mu}},(j=0,1),
\end{equation}
if
\begin{equation}
\mathbf{X}_{0}^{[1]}L_{0}+(D_{s}{\xi}_{0})L_{0}=D_{s}A_{0},
\label{12}
\end{equation}
and
\begin{equation}
\mathbf{X}_{1}^{[1]}L_{0}+\mathbf{X}_{0}^{[1]}L_{1}+(D_{s}{\xi}_{1})L_{0}
+(D_{s}{\xi}_{0})L_{1}=D_{s}A_{1}. \label{13}
\end{equation}
Here ${L}_{0}$ is the exact Lagrangian corresponding to the exact
equations and ${L}_{0}+\epsilon{L}_{1}$ the first-order perturbed
Lagrangian corresponding to the first-order perturbed equations. The
perturbed equations (3) and (16) always have the approximate
symmetry generators $\epsilon \mathbf{X}_{0}$ which are known as
trivial approximate symmetries and $\mathbf{X}$ given by (1) and
(12) with $\mathbf{X}_{0}\neq 0$ is called a non-trivial approximate
symmetry. These approximate symmetries of a manifold, form an
approximate Lie algebra \cite{Gaz}.

The first-order approximate first integrals are defined by setting
$I=I_0+\epsilon I_1$, $L=L_0+\epsilon L_1$, $\xi=\xi_0+\epsilon
\xi_1$, $\eta=\eta_0+\epsilon \eta_1$ and $A=A_0+\epsilon A_1$ in
(9) and equating the coefficients of like powers of $\epsilon$ on
both sides. This gives the zeroth (exact part) and first-order
approximate part of the first-order approximate first integrals
\begin{align}
I_0&={\xi_0}L_0+(\eta^{\mu}_0-\dot{x}^{\mu}{\xi_0})\frac{\partial
L_0}{\partial \dot{x}^{\mu}}-A_0,\\
I_1&={\xi_0}L_1+{\xi_1}L_0+(\eta^{\mu}_0-\dot{x}^{\mu}{\xi_0})\frac{\partial
L_1}{\partial \dot{x}^{\mu}}+
(\eta^{\mu}_1-\dot{x}^{\mu}{\xi_1})\frac{\partial L_0}{\partial
\dot{x}^{\mu}}-A_1.
\end{align}
If $I_0$ vanishes, then $I$ is called an unstable approximate first
integral and is otherwise called stable. A detailed discussion on
the approximate first integrals for Hamiltonian dynamical systems is
given in \cite{GU}.

The Weyl tensor in component form is given by
\begin{equation}
C^a_{\;\;bcd}=R^a_{\;\;bcd}-\frac{1}{2}(\delta^a_cR_{bd}-\delta^a_dR_{bc}+g_{bd}R^a_{\;\;c}
-g_{bc}R^a_{\;\;d})+\frac{1}{6}R(\delta^a_dg_{bc}-\delta^a_cg_{bd}).
\end{equation}
Here $R^a_{\;\;bcd}$ is the Riemann curvature tensor, $R_{ab}$ is
the Ricci tensor, $R$ is the Ricci scalar, $g_{ab}$ is the metric
tensor and $\delta^a_b$ is the Kroneker delta. For a 4 dimensional
spacetime the Weyl tensor has 10 independent components
\cite{ESEFEs}. If the Weyl tensor vanishes in a neighborhood of a
spacetime, the neighborhood is locally conformally equivalent to the
Minkowski spacetime. Thus the Weyl tensor has geometric meaning
independent of any physical interpretation.

The stress-energy tensor can be calculated from the EFEs
\begin{equation}
T_{ab}=\frac{1}{\kappa }(R_{ab}-\frac{1}{2}Rg_{ab}),
\end{equation}
where $\kappa$ is the gravitational coupling. For a 4-dimensional
spacetime this tensor has 10 independent components. At each event
of the spacetime this tensor gives the energy density, momentum
density and stress as measured by any and all observers at that
event. Since for gravitational wave spacetimes $T_{ab}$ is always
zero and $C^a_{\;\;bcd}$ is nonzero, there is no matter or energy or
momentum however there is the Weyl curvature. If there is no mass or
energy at a given event, the Ricci tensor vanishes. If it were not
for the Weyl tensor, this would mean that matter at one place could
not have gravitational influence on distant matter separated by a
void. Thus the Weyl tensor represents that part of spacetime
curvature which can propagate across and curve up a void.

\section{Second-order approximate symmetries and energy rescaling: plane
wave spacetimes}

The line element for pp-waves \cite{MM}, is
\begin{eqnarray}
ds^{2}=h\omega^2[(x^2-y^2)\sin(\omega(t-z))+2xy\cos(\omega(t-z))]
(dt^{2}+dz^2-2dtdz)+dt^2\nonumber\\-dx^{2}-dy^2-dz^2,
\end{eqnarray}
where $h$ is the amplitude of the wave and $\omega$ is the
frequency.

Now we remove the $t$-dependent part of (21) and putting $h=1$ to
define a static spacetime
\begin{equation}
ds^{2}=\omega^2[(x^2-y^2)+2xy](dt^{2}+dz^2-2dtdz)+dt^2-dx^{2}-dy^2
-dz^2.
\end{equation}
To obtain the approximate symmetries of pp-waves the exact pp-waves
are considered as a perturbation on the static spacetime (22). For
this purpose the amplitude $h=\epsilon$, is taken as a small
parameter and the line element of the perturbed pp-waves is
\begin{eqnarray}
ds^{2}=\omega^2[x^2-y^2+2xy+\epsilon
\{(x^2-y^2)\sin(\omega(t-z))+2xy\cos(\omega(t-z))\}](dt^{2}\nonumber\\+dz^{2}
-2dtdz)+dt^2-dx^2-dy^2-dz^2.
\end{eqnarray}
For this perturbed pp-wave spacetime (23) we have the system of
first-order perturbed geodesic equations and there do not appear
$\epsilon^2$,
\begin{align}
\ddot{t}+\omega^2(\dot{t}-\dot{z})\{(x+y)\dot{x}+(x-y)\dot{y}\}+\epsilon
[\frac{\omega^3}{2}\{(x^2-y^2)\cos\omega(z-t)\nonumber\\+2xy\sin\omega(z
-t)\}(\dot{t}^2+\dot{z}^2-\dot{t}\dot{z})-\omega^2(\dot{t}-\dot{z})\{x\sin\omega(z
-t)\nonumber\\-y\cos\omega(z-t)\}\dot{x}+\omega^2\{y\sin\omega(z-t)+x\cos\omega(z-t)\}
\dot{y}]=0,\\
\ddot{x}+[\omega^2(x+y)-\epsilon\omega^2\{
x\sin\omega(z-t)-y\cos\omega(z-t)\}](\dot{t}^2+\dot{z}^2-\dot{t}\dot{z})=0,\\
\ddot{y}+[\omega^2(x+y)-\epsilon\omega^2\{
x\cos\omega(z-t)-y\sin\omega(z-t)\}](\dot{t}^2+\dot{z}^2-\dot{t}\dot{z})=0,\\
\ddot{z}+\omega^2(\dot{t}-\dot{z})\{(x+y)\dot{x}+(x-y)\dot{y}\}+\epsilon
[\frac{\omega^3}{2}\{(x^2-y^2)\cos\omega(z-t)\nonumber\\+2xy\sin\omega(z
-t)\}(\dot{t}^2+\dot{z}^2-\dot{t}\dot{z})-\omega^2(\dot{t}-\dot{z})\{x\sin\omega
(z-t)\nonumber\\-y\cos\omega(z-t)\}
\dot{x}+\omega^2\{y\sin\omega(z-t)+x\cos\omega(z-t)\} \dot{y}]=0.
\end{align}
Since there is no quadratic term in $\epsilon$, in the above
geodesic equations, we cannot apply the definition of second-order
approximate symmetries, which gives us the energy rescaling factor
to them. This behavior is consistent with the pp-wave geometry in
which the wave front moves as parallel planes and the spacetime
curvature is zero before and after the pp-wave pulse \cite{MTW}.

To obtain a better understanding of the energy rescaling in plane
gravitational waves we apply the definition of second-order
approximate symmetries of ODEs to the second-order perturbed
geodesic equations (29) - (32) for the artificially constructed
example of plane symmetric wave-like spacetime \cite{IA}. For this
purpose a non-flat plane symmetric static spacetime \cite{TQZ}, was
considered with $\mu(x)=\nu^2(x)=(x/X)^2$,
\begin{equation}
ds^{2}= e^{2\nu(x)}dt^2-dx^2-e^{2\mu(x)}(dy^2+dz^2),
\end{equation}
where $X$ is a constant having the same dimensions as $x$. Since
gravitational waves are non-static spacetimes therefore the static
spacetime (28) was perturbed with a time-dependent small parameter
(for definiteness by $\epsilon$$t$) to make it slightly non-static.
For this the metric (28) was taken with $\nu(x)=(x/X+\epsilon t/T)$
and $\mu(x)=(x^2/X^2+\epsilon t/T)$, where $T$ is a constant having
dimensions of $t$. Retaining $\epsilon^2$ and neglecting its higher
powers second-order perturbed geodesic equations are obtained
\begin{align}
\ddot{t}+\frac{2}{X}\dot{t}\dot{x}-\frac{\epsilon}{T}[\dot{t}^2-(\dot{y}^2+\dot{z}^2)e^{2((x/X)^2-x/X)}]
+\frac{t\epsilon^2}{T^2}[\dot{t}^2+(\dot{y}^2+\nonumber\\\dot{z}^2)
e^{2((x/X)^2-x/X)}]+O(\epsilon^3)&=0,\\
\ddot{x}+\frac{\dot{t}^2}{X}e^{2x/X}-\frac{2x}{X^2}(\dot{y}^2+\dot{z}^2)e^{2(x/X)^2}
+\frac{2t\epsilon}{TX}[\dot{t}^2e^{2x/X}-\frac{2x}{X}(\dot{y}^2\nonumber\\+\dot{z}^2)e^{2(x/X)^2}]
+\frac{t^2\epsilon^2}{XT^2}[\dot{t}^2e^{2x/X}
-\frac{2x}{X}(\dot{y}^2+\dot{z}^2)e^{2(x/X)^2}]+O(\epsilon^3)&=0,\\
\ddot{y}+\frac{4x}{X^2}\dot{x}\dot{y}+\frac{2\epsilon}{T}\dot{t}\dot{y}
-\frac{2t\epsilon^2}{T^2}\dot{t}\dot{y}+O(\epsilon^3)&=0,\\
\ddot{z}+\frac{4x}{X^2}\dot{x}\dot{z}+\frac{2\epsilon}{T}\dot{t}\dot{z}
-\frac{2t\epsilon^2}{T^2}\dot{t}\dot{z}+O(\epsilon^3)&=0.
\end{align}

The Lie symmetry algebra of the exact or unperturbed geodesic
equations (i.e. when $\epsilon=0$, in (29) - (32)) includes the
generators of the dilation algebra ${\partial}/{\partial s}$,
$s{\partial}/{\partial s}$, corresponding to
\begin{equation}
\xi(s)=c_0s+c_1.
\end{equation}
In the determining equations for the first-order approximate
symmetries of the geodesic equations for the Schwarzschild spacetime
\cite{KMQ} the terms involving $\xi_s=c_0$ cancel out. Here the
terms involving $\xi_s=c_0$ do not cancel automatically but, like
the RN \cite{IMQ} and charged-Kerr \cite{IMQ1} spacetimes, collect a
scaling factor to cancel out. In this case the scaling factor is
\begin{equation}
\frac{t}{T^2}[\dot{t}^2+(\dot{y}^2+\dot{z}^2) e^{2((x/X)^2-x/X)}].
\end{equation}

Energy conservation is related with time translation and $\xi$ is
the coefficient of ${\partial}/{\partial s}$ in the point
transformation given by (5), where $s$ is the proper time. The
scaling factor (34) corresponds to the rescaling of energy of a test
particle in this wave-like spacetime field. Since the scaling factor
(34) involve the derivatives of the coordinates and the derivatives
only apply to the paths of the particles. To get energy in the
spacetime field the derivatives of the coordinates are replaced by
the first integrals. Therefore we get
\begin{equation}
\frac{t}{4T^2}[e^{-4x/X}+2e^{-2(x/X)(x/X+1)}].
\end{equation}
This energy expression is plotted below for different values of $t$
and $x$, using Mathematica 5.0. The values of $X$ and $T$ are
arbitrary. The above scaling factor for this wave-like spacetime
depends linearly on $t$ and in both diagrams below the energy in the
gravitational field increases linearly with time. In Fig. 1 the
energy is seen to decrease along $x$ and disappear sharply close to
$x=0$. To see the variation with $x$ we enlarge the diagram by
reducing the range of $x$ in Fig. 2. As we move along $x$ the
increase in energy with time becomes gradual. Since the small
parameter $\epsilon$, (which is considered as the strength of the
wave) is arbitrary the units of energy are arbitrarily chosen.
Throughout this paper gravitational units are used and space, time
and mass are given in seconds.

\begin{figure}
\begin{center}
\includegraphics[width=4in, height=3in]{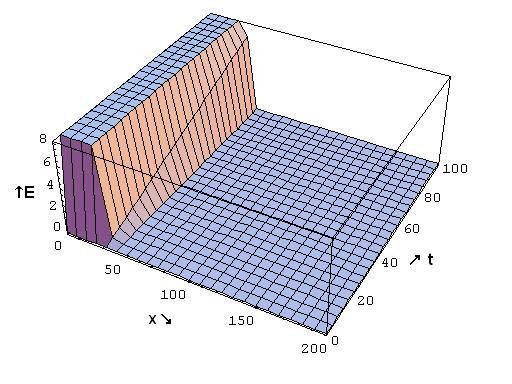}\\
\caption{\it Plane symmetric gravitational wave-like spacetime. The
energy increases indefinitely in time close to $x=0$ and then
disappears suddenly after some distance. The small parameter
$\epsilon$, (considered as strength of the wave) is arbitrary in all
the spacetimes discussed in this paper. Thus the units of energy are
chosen arbitrarily. Throughout this paper gravitational units are
adopted and space, time and mass are given in seconds.}\label{}
\includegraphics[width=4in, height=3in]{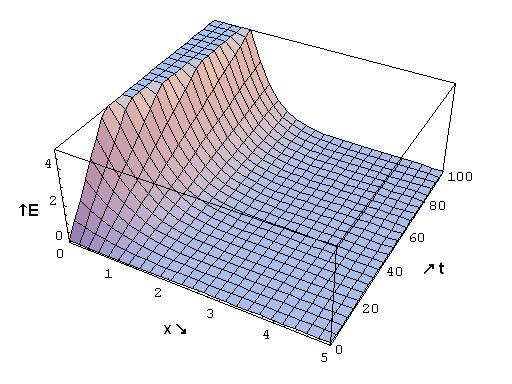}\\
\caption{\it This is an expanded version of Fig. 1. Here the range
of $x$ is shrunk and it is seen that the energy decreases smoothly
with distance.}\label{}
\end{center}
\end{figure}

\pagebreak

Though the Weyl tensor gives information about the gravitational
energy of the spacetime, it is not clear how to obtain a {\it measure}
of the energy in it. For the pure gravitational part of the perturbed
pp-wave spacetime the independent nonzero components of the Weyl
tensor are
\begin{align}
C^0_{\;\;101}=-\omega^2[\omega^2(x^2-y^2-2xy)+1+\epsilon\{2\omega^2(x^2-y^2-xy)
\sin\omega(z-t)-2\omega^2xy\nonumber\\\cos\omega(z-t)-\sin\omega(z-t)\}]+O(\epsilon^2)
=C^0_{\;\;113}=C^1_{\;\;313}=-C^0_{\;\;202}=C^0_{\;\;223}=C^2_{\;\;323},\nonumber\\
C^0_{\;\;102}=-\omega^2[\omega^2(y^2-x^2-2xy)+1+\epsilon\{\omega^2(x^2-y^2)
\sin\omega(z-t)-\omega^2(x^2-y^2+\nonumber\\4xy)\cos\omega(z-t)+\cos\omega(z-t)\}]+O(\epsilon^2)
=C^0_{\;\;123}=C^0_{\;\;213}=C^1_{\;\;323}.
\end{align}

In the literature \cite{GSH} the Weyl tensor is usually defined with
valence (1, 3). In spinors it is naturally given as a tensor of valence
(0, 4) \cite{RP1}. For usual purposes the form does not matter, but for
differential symmetries of the tensor the form is crucial \cite{AK}.
In covariant form the components of the Weyl tensor are
\begin{align}
C_{0101}=-\omega^2[1-\epsilon \sin\omega(z-t)]
=C_{0113}=C_{1313}=-C_{0202}=C_{0223}=C_{2323},\nonumber\\
C_{0102}=-\omega^2[1+\epsilon \sin\omega(z-t)]
=C_{0123}=C_{0213}=C_{1323}.
\end{align}
From here it appears that the (0, 4) form may give the physically
relevant quantities as the space dependence in (36) does not seem to
correspond to the geometry of the pp-wave, while (37) does. Here the
pure gravitational field which ``curves up the void'' is seem to be
sinusoidal. For this spacetime there is obviously no nonzero
component of the stress-energy tensor.

There are 6 nonzero components (up to first-order in $\epsilon$) of
the Weyl tensor for the above plane symmetric wave-like spacetime
\begin{align}
C^0_{\;\;101}&=\frac{1}{3X^3}(2x+X)+O(\epsilon^2),\nonumber\quad\\
C^0_{\;\;202}&=C^0_{\;\;303}=\frac{e^{2x^2/X^2}}{6X^3}(1+\epsilon\frac{2t}{T})
(2x+X)+O(\epsilon^2),\nonumber\quad\\
C^1_{\;\;212}&=C^1_{\;\;313}=-C^0_{\;\;202},\quad
C^2_{\;\;323}=-2C^0_{\;\;202}.
\end{align}
From Figs. 1 and 2 (where the wave is along the $x$ direction), it is
clear that the energy in the gravitational field of the plane
wave-like spacetime increases with time. Therefore the first
component of the Weyl tensor must depend on $t$ linearly which
corresponds to the covariant form (given below) and not the mixed
form.
\begin{align}
C_{0101}&=\frac{e^{2x/X}}{3X^3}(1+\epsilon\frac{2t}{T})(2x+X)+O(\epsilon^2),\nonumber\quad\\
C_{0202}&=C_{0303}=\frac{e^{{2x^2/X^2}+{2x/X}}}{6X^3}(1+\epsilon\frac{4t}{T})
(2x+X)+O(\epsilon^2),\nonumber\quad\\
C_{1212}&=C_{1313}=-C_{0202},\quad C_{2323}=-2C_{0202}.
\end{align}
The nonzero components of the stress-energy tensor for this
wave-like spacetime are
\begin{align}
T_{00}&=\frac{4e^{2x/X}}{\kappa X^4}(1+\epsilon
\frac{2t}{T})(1+\frac{3x^2}{X^2})+O(\epsilon^2), \quad
T_{11}=\frac{-8x}{\kappa
X^4}(x+X)+O(\epsilon^2),\nonumber\\
T_{22}&=T_{33}=\frac{e^{2x^2/X^2}}{\kappa
X^4}(1+\epsilon\frac{2t}{T})(2xX+4x^2+3X^2)+O(\epsilon^2),\quad\nonumber\\
T_{01}&=\frac{\epsilon}{\kappa TX^2}(x+X)+O(\epsilon^2).
\end{align}
It is worth noting that the $x$-direction stress has no approximate
part of first-order and the approximate part of the energy increases
linearly with time and quadratically at large distances. More
interestingly there is an approximate momentum in the $x$-direction
that increases linearly with the value of $x$. This linear increase
in energy was built into the metric and it entails the momentum in
the $x$-direction.

We give the ratio of energy density imparted to the matter field
\begin{equation}
E_{imp}=\frac{(T_{00})_P}{(T_{00})_E},
\end{equation}
where $(T_{00})_E$ and $(T_{00})_P$ are the energy densities of the
exact (i.e. when $\epsilon=0$) and first-order approximate
spacetimes respectively. For the plane symmetric wave-like spacetime
we have
\begin{align}
E_{imp}=\epsilon\frac{2t}{T}.
\end{align}

\section{Approximate Noether symmetries of plane wave spacetimes}

The Lagrangian defined for (28) is \cite{IA} (throughout this paper
the Lagrangian of a spacetime would means the Lagrangian for the
geodesic equations of the spacetime)
\begin{equation}
L= e^{2x/X}\dot{t}^2-\dot{x}^2-e^{2x^2/X^2}(\dot{y}^2+\dot{z}^2).
\end{equation}
Its symmetry generators are
\begin{equation}
\quad \mathbf{X}_{0}=\frac{\partial}{\partial t},\quad
\mathbf{X}_{1}=\frac{\partial}{\partial y},\quad
\mathbf{X}_{2}=\frac{\partial}{\partial z},\quad
\mathbf{X}_{3}=y\frac{\partial}{\partial
z}-z\frac{\partial}{\partial y},\quad
\mathbf{Y}_{0}=\frac{\partial}{\partial s},\quad A=c ,
\end{equation}
where $c$ is a constant, $\mathbf{X}_{0}$ corresponds to energy
conservation, $\mathbf{X}_{1}$ and $\mathbf{X}_{2}$ correspond to
linear momentum conservation along $y$ and $z$, while
$\mathbf{X}_{3}$ corresponds to angular momentum conservation in the
$yz$ plane \cite{AQs}.

The first-order perturbed Lagrangian is
\begin{equation}
\quad
L=e^{2x/X}\dot{t}^2-\dot{x}^2-e^{2x^2/X^2}(\dot{y}^2+\dot{z}^2)+
\frac{2\epsilon
t}{T}[e^{2x/X}\dot{t}^2-e^{2x^2/X^2}(\dot{y}^2+\dot{z}^2)]
+O(\epsilon^2),
\end{equation}
yielding the non-trivial approximate symmetry
\begin{equation}
\mathbf{X}_a= \frac{\partial}{\partial t}-\epsilon \frac{1}{T}(t
\frac{\partial}{\partial t} + y \frac{\partial}{\partial y}
+z\frac{\partial}{\partial z}),
\end{equation}
along with the trivial symmetries, and the gauge function $A_{1}$ is
again a constant. The stable first integral for the symmetry
generator (46) is
\begin{equation}
I=2e^{2x/X}\dot{t}+\frac{2\epsilon}{T}[e^{2x/X}t\dot{t}+e^{2(x/X)^2}(y\dot{y}+z\dot{z})].
\end{equation}
Contract the energy momentum vector with the time-like approximate
Noether symmetry generator (46), to obtain the conserved quantity
\begin{equation}
Q=E-\frac{\epsilon}{T}(t\dot{E}+y\dot{p_y}+z\dot{p_z}),
\end{equation}
where $E$ is the energy and $p$ is the momentum. This gives the
energy non-conservation due to time variation. That is the energy
imparted to the test particles with energy and momentum given by
(48). However this does not give the energy in the spacetime field.

To check the conserved quantities in the pp-wave spacetime, we
investigate the first-order approximate Noether symmetries for this
spacetime. The Lagrangian for the pp-wave spacetime (21) is
\begin{equation}
L=h\omega^2[(x^2-y^2)\sin(\omega(t-z))+2xy\cos(\omega(t-z))](\dot{t}^{2}+\dot{z}^2
-2\dot{t}\dot{z})+\dot{t}^2-\dot{x}^{2}-\dot{y}^2- \dot{z}^2.
\end{equation}
This Lagrangian admits the following symmetry generators
\begin{equation}
\mathbf{X}_{0}=\frac{\partial}{\partial t}+\frac{\partial}{\partial
z},\quad\mathbf{Y}_{0}=\frac{\partial}{\partial s}\quad {\rm
and}\quad A=c\quad (\rm constant).
\end{equation}

The Lagrangian for the static spacetime (22)
\begin{equation}
L=\omega^2[(x^2-y^2)+2xy](\dot{t}^{2}+\dot{z}^2
-2\dot{t}\dot{z})+\dot{t}^2-\dot{x}^{2}-\dot{y}^2- \dot{z}^2,
\end{equation}
has 3 symmetry generators
\begin{equation}
\mathbf{X}_{0}=\frac{\partial}{\partial t},\quad
\mathbf{X}_{1}=\frac{\partial}{\partial z},\quad
\mathbf{Y}_{0}=\frac{\partial}{\partial s},
\end{equation}
and the gauge function is a constant.

The Lagrangian for the perturbed pp-wave spacetime (23) is
\begin{align}
L=\omega^2[x^2-y^2+2xy+\epsilon
\{(x^2-y^2)\sin(\omega(t-z))+2xy\cos(\omega(t-z))\}](\dot{t}^{2}+\dot{z}^2\nonumber\\
-2\dot{t}\dot{z})+\dot{t}^2-\dot{x}^{2}-\dot{y}^2- \dot{z}^2.
\end{align}
For $\epsilon=0$, the above Lagrangian (53) reduces to (51). Using
this first-order perturbed Lagrangian and the three exact symmetry
generators given by (52) in (16), in the resulting system of
determining equations two constants corresponding to the exact
symmetry generators appears. These two generators have to be
eliminated for consistency of the determining equations, making them
homogeneous. The resulting system is the same as for the static
(exact) spacetime (23). Thus there is no non-trivial approximate
symmetry for this perturbed Lagrangian and the gauge function is a
constant. Hence we can not obtain the conserved quantity in the case
of perturbed pp-wave spacetime. Only the three exact symmetry
generators are recovered as trivial first-order approximate Noether
symmetries which gives trivial first-order approximate conservation
laws for energy and linear momentum along $z$.

\section{Second-order approximate symmetries and energy rescaling:
cylindrical wave spacetimes}

The line element of the cylindrically symmetric exact waves
\cite{ER} is
\begin{equation}
ds^{2}=e^{2(\gamma -\psi
)}(dt^{2}-d\rho^2)-\rho^2e^{-2\psi}d\phi^2-e^{2\psi}dz^2,
\end{equation}
where $\gamma$ and $\psi$ are arbitrary functions of $t$ and $\rho$,
subject to the vacuum EFEs
\begin{equation}
\psi''+\frac{1}{\rho}\psi'-\ddot{\psi}=0,\quad\gamma'=\rho(\psi'^2+\dot{\psi}^2),
\quad\dot{\gamma}=2\rho\dot{\psi}\psi',
\end{equation}
where dot denotes differentiation with respect to $ t$ and prime
with respect to $\rho$. The solution of (55) is given by \cite{WW1}
\begin{align}
\psi&=AJ_0(\omega \rho)\cos(\omega t)+BY_0(\omega \rho)\sin(\omega t),\quad\\
\gamma&=\frac{1}{2}\omega\rho[(A^2J_0{J_0}'-B^2Y_0{Y_0}')\cos(2\omega
t) -AB\{(J_0{Y_0}'+Y_0{J_0}')\sin(2\omega
t)\nonumber\\&-2(J_0{Y_0}'-Y_0{J_0}')\omega t\}].
\end{align}
This metric has two KVs ${\partial}/{\partial \phi}$ and
${\partial}/{\partial z}$ \cite{ESEFEs}; this means that there is
only azimuthal angular momentum conservation and linear momentum
conservation along $z$.

To discuss the approximate symmetries of cylindrical waves first a
static spacetime is defined as follows. We remove the $t$-dependent
part in (54) and put the strength of the wave $A=1$ and $B=0$. $Y_0$
become singular at the origin and putting $B=0$ give us rid of this
singular behavior.
\begin{equation}
ds^{2}=e^{2(\gamma_0 -\psi_0
)}(dt^{2}-d\rho^2)-\rho^2e^{-2\psi_0}d\phi^2-e^{2\psi_0}dz^2,
\end{equation}
where
\begin{equation}
\psi_0=J_0(\omega\rho),\quad
\gamma_0=\frac{\omega\rho}{2}J_0(\omega\rho)J'_0(\omega\rho).
\end{equation}
For the approximate case we put the strength of the wave as a small
parameter i.e. $A=\epsilon$ and take the exact wave as a
perturbation on the static metric (58) in the following way.
\begin{align}
\psi&=J_0(\omega\rho)(1+\epsilon\cos(\omega
t))=\psi_0+\epsilon\psi_1\nonumber\\
\gamma&=\frac{\omega\rho}{2}J_0(\omega\rho)J'_0(\omega\rho)(1+\epsilon^2\cos(2\omega
t))=\gamma_0+\epsilon^2\gamma_1.
\end{align}
Thus the second-order perturbed geodesic equations are
\begin{align}
\ddot{t}+2(\gamma_0'-\psi_0')\dot{t}\dot{\rho}-\epsilon[\dot{\psi_1}(\dot{t^2}
+\dot{\rho^2})+\rho^2e^{-2\gamma_0}\dot{\psi_1}\dot{\phi^2}-e^{2(2\psi_0
-\gamma_0)}\dot{\psi_1}\dot{z^2}+2{\psi_1'}\dot{t}\dot{\rho}]+\epsilon^2
[\dot{\gamma_1}\nonumber\\(\dot{t^2}+\dot{\rho^2})+4e^{2(2\psi_0-\gamma_0)}
\psi_1\dot{\psi_1}\dot{z^2}-2\gamma_1'\dot{t}\dot{\rho}]+O(\epsilon^3)=0,\\
\ddot{\rho}+(\gamma_0'-\psi_0')(\dot{t^2}+\dot{\rho^2}+\dot{t}\dot{\rho})
+\rho\e^{-2\gamma_0}(\psi_0'-1)\dot{\phi}^2+e^{2(2\psi_0-\gamma_0)}\psi_0'\dot{z}^2
-\epsilon[\psi_1'(\dot{t^2}+\dot{\rho^2})\nonumber\\-\rho^2e^{-2\gamma_0}\dot{\psi_1}\dot{\phi}^2
-e^{2(2\psi_0-\gamma_0)}(4\psi_0'\psi_1-\psi_1')\dot{z}^2+\psi_1'\dot{t}\dot{\rho}]
+\epsilon^2[\gamma_1'(\dot{t^2}+\dot{\rho^2})-\rho\gamma_1e^{-2\gamma_0}(\psi_0'\nonumber\\-1)
\dot{\phi}^2+2e^{2(2\psi_0-\gamma_0)}(4\psi_0'\psi_1^2-\gamma_1\psi_0'-\psi_1\psi_1')
\dot{z^2}+\gamma_1'\dot{t}\dot{\rho}]+O(\epsilon^3)=0,\\
\ddot{\phi}+\frac{1}{\rho}(1-\psi_0')\dot{\rho}\dot{\phi}-\epsilon
(\psi_1'\dot{\rho}+\dot{\psi_1}\dot{t})\dot{\phi}-\frac{\epsilon^2}{\rho}
(1-\psi_0')\dot{\rho}\dot{\phi}+O(\epsilon^3)=0,\\
\ddot{z}+\psi_0'\dot{\rho}\dot{z}+\epsilon(\psi_1'\dot{\rho}+\dot{\psi_1}\dot{t})\dot{z}
-2\epsilon^2\psi_0'\psi_1^2\dot{\rho}\dot{z}+O(\epsilon^3)=0.
\end{align}
The dot and prime over $\gamma_0$, $\psi_0$, $\gamma_1$ and $\psi_1$
denote derivatives with respect to $\omega t$ and $\omega\rho$
respectively. For this perturbed wave spacetime the scaling factor
is
\begin{equation}
\dot{\gamma_1}(\dot{t^2}+\dot{\rho^2})+4e^{2(2\psi_0-\gamma_0)}
\psi_1\dot{\psi_1}\dot{z^2}-2\gamma_1'\dot{t}\dot{\rho}.
\end{equation}
To replace the derivative of the coordinates $t$, $z$ and $\rho$ the
exact first integrals and the metric (58) are used. Further it is
assumed that there is no initial velocity in the $z$ and $\phi$
directions. Hence $\dot{z}$ and $\dot{\phi}$ vanishes and the
following scaling factor is obtained
\begin{equation}
\dot{\gamma_1}e^{2(\psi_0-\gamma_0)}[e^{2(\psi_0-\gamma_0)}+e^{3(\psi_0-\gamma_0)}
-1]-2\gamma_1'e^{3(\psi_0-\gamma_0)}(e^{3(\psi_0-\gamma_0)}-1)^{1/2},
\end{equation}
where $\gamma_1$ is given in (60). This scaling factor involves the
Bessel function of first kind and its derivatives. The asymptotic
representation of the Bessel function of first kind for large value
of the argument is given in \cite{NRS}. Using this asymptotic
representation of the Bessel function in (66), we obtain an
asymptotic representation of it as follows
\begin{equation}
\frac{3\times2^{11/4}}{\pi^{3/2}}[(|\cos(\omega\rho)|)^{3/2}\sin(2\omega
t)](\omega\rho)^{-1/2}+O([\omega\rho]^{-3/2}).
\end{equation}
Thus the energy in this perturbed spacetime field is rescaled by the
factor (67). It is plotted below for different values of $t$, $\rho$
and $\omega$ (in radians per second), in which the energy oscillates
between positive and negative values and goes to zero as $\rho$
tends to infinity. Here the behavior is much more recognizably
wave-like. Since the strength of the wave, $A = \epsilon$, is
arbitrary the energy is given in arbitrarily chosen units.

\begin{figure}
\begin{center}
\includegraphics[width=4in, height=3in]{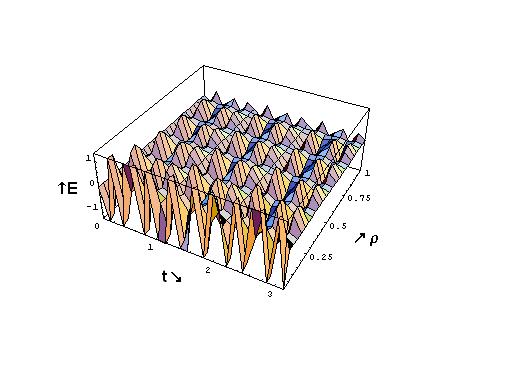}\\
\caption{\it Cylindrically symmetric gravitational waves with
$\omega=15$. The gravitational energy oscillates between positive
and negative values and disappears as $\rho$ approaches very large
value. The units of energy are arbitrary in all diagrams.}\label{}
\includegraphics[width=4in, height=3in]{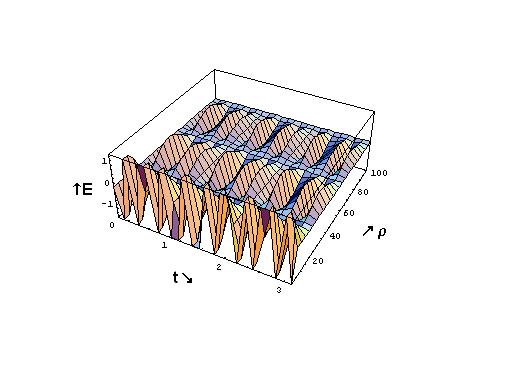}\\
\caption{\it To see the behavior of energy for comparatively larger
distance, therefore the range of $\rho$ is extended to 100
units.}\label{}
\end{center}
\end{figure}

\begin{figure}
\begin{center}
\includegraphics[width=4in, height=3in]{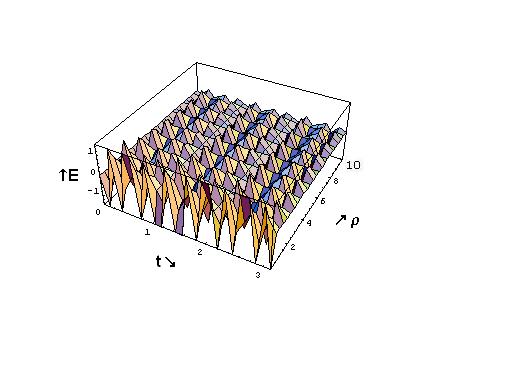}\\
\caption{\it To see a further extended version of the above fig. 4,
the rang of $\rho$ is given in units of $10^{5}$.}\label{}
\end{center}
\end{figure}

\begin{figure}
\begin{center}
\includegraphics[width=4in, height=3in]{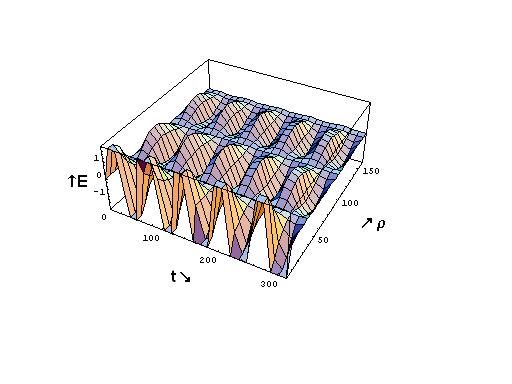}\\
\caption{\it Here the value of the frequency is comparatively small
i.e. $\omega=0.05$. To see the variation along time, therefore the
rang of $t$ is kept larger.}\label{}
\end{center}
\end{figure}

\pagebreak

For completeness we also investigate second-order approximate
symmetries of the geodesic equations for the cylindrically symmetric
wave-like spacetime. For this purpose a cylindrically symmetric
static metric is taken \cite{QZ}
\begin{equation}
ds^{2}= e^{\nu(\rho )}dt^2-d\rho^{2}-e^{\mu(\rho)}(a^2d\phi^2+dz^2),
\end{equation}
with $\nu(\rho)=(\rho/R)^2$ and $\mu(\rho)=(\rho/R)^3$, where $R$ is
a constant having the same dimensions as $\rho$. For the approximate
symmetries of this cylindrical wave-like spacetime,
$\nu(x)=(\rho/R)^2+2\epsilon t/T$ and $\mu(x)=(\rho/R)^3+2\epsilon
t/T$ are taken in the metric (68), where $T$ is a constant having
dimensions of $t$. Like the plane symmetric case the scaling factor
for this cylindrical wave-like spacetime is
\begin{equation}
\frac{t}{4T^2}[e^{-2(\rho/R)^2}+2e^{-(\rho/R)^2(\rho/R+1)}].
\end{equation}
Thus the energy in the field of this wave-like spacetimes is
rescaled by the factor (69). The plots for this case, are similar to
those for the plane wave-like case discussed in section 3, with $x$
replaced by $\rho$.

The nonzero components of the Weyl tensor for the cylindrically
symmetric perturbed wave spacetime are
\begin{eqnarray}
C^0_{\;\;101}=-\frac{1}{3}[\psi_0''-\gamma_0''+2\psi_0'^2-2\psi_0'
+\epsilon(4\psi_0'\psi_1'+\psi_1''-\ddot{\psi_1}-2\psi_1'
+4\psi_0'\psi_1\nonumber\\-4\psi_0'^2\psi_1-2\psi_0''\psi_1+2\gamma_0''\psi_1)]
+O(\epsilon^2),\nonumber \\
C^0_{\;\;202}=\frac{\rho^2}{6}[4\psi_0''-\gamma_0''+8\psi_0'^2+
3\gamma_0'-2\psi_0'-6\psi_0'\gamma_0'+2\epsilon(2\psi_1''+\ddot{\psi_1}
+8\psi_0'\psi_1'\nonumber\\-\psi_1'-3\gamma_0'\psi_1')]
+O(\epsilon^2),\nonumber \\
C^0_{\;\;303}=\frac{e^{2(2\psi_0-\gamma_0)}\rho}{6}[2\psi_0''+\gamma_0''+4\psi_0'^2+
3\gamma_0'+2\psi_0'-6\psi_0'\gamma_0'+2\epsilon(\psi_1''+2\ddot{\psi_1}
\nonumber\\+4\psi_0'\psi_1'+\psi_1'-3\gamma_0'\psi_1')]
+O(\epsilon^2),\nonumber\\
C^1_{\;\;212}=-e^{-4\psi_0}C^0_{\;\;303},\quad
C^1_{\;\;313}=\frac{1}{\rho}e^{2(2\psi_0-\gamma_0)}C^0_{\;\;202}\nonumber\\
C^2_{\;\;323}=\frac{e^{2(2\psi_0-\gamma_0)}}{3}[\gamma_0''-\psi_0''
+\psi_0'-2\psi_0'^2-2\epsilon(\psi_1''-\ddot{\psi_1}+2\psi_0'\psi_1'
-\frac{1}{\rho}\psi_1')]+O(\epsilon^2),\nonumber \\
C^0_{\;\;212}=-\epsilon\rho^2e^{-2\gamma_0}[(2\psi_0'+\gamma_0')\dot{\psi_1}+
\dot{\psi_1}']+O(\epsilon^2),\nonumber\\
C^0_{\;\;313}=\epsilon
e^{2(2\psi_0-\gamma_0)}[(2\psi_0'-\gamma_0')\dot{\psi_1}+
\dot{\psi_1}']+O(\epsilon^2).
\end{eqnarray}
This yields the pure gravitational field for this cylindrically
symmetric perturbed wave spacetime. As it is evident from Figs. 3 to
6, that the energy in the gravitational field oscillates and then
vanishes for large $\rho$, here all the components of the Weyl
tensor also depend on the Bessel function of the first kind and its
derivatives which oscillates and goes to zero as $\rho$ approaches
very large value. The last two components only appear for the
approximate part of the spacetime. In this case the components of
the Weyl tensor in the covariant form are not very different from
those in the mixed form given above and therefore we do not give
them separately.

The non-vanishing components of the stress-energy tensor are
\begin{align}
T_{00}=T_{11}=\kappa[\psi_0'^2-\gamma_0'+2\epsilon\psi_0'\psi_1']
+O(\epsilon^2),\nonumber \\
T_{22}=\kappa\rho^2e^{2(\gamma_0-2\psi_0)}
[\psi_0'^2+\gamma_0''+2\epsilon\psi_0'\psi_1']+O(\epsilon^2),\nonumber \\
T_{33}=\kappa
e^{-2(\gamma_0-2\psi_0)}[2\psi_0''-\gamma_0''-\psi_0'^2+2\psi_0'
+\frac{2}{\rho}\epsilon\{\psi_1''-\ddot{\psi_0}+2\psi_1\nonumber\\
(\psi_0''-\gamma_0''-\psi_0'^2 +3\psi_0'-\psi_0'\psi_1'+2
(\psi_1'-2\psi_0'\psi_1))\}]+O(\epsilon^2),\nonumber \\
T_{01}=\kappa\epsilon\psi_0'\dot{\psi_1}+O(\epsilon^2).
\end{align}
Like the components of the Weyl tensor the above components of the
stress-energy tensor also depend on the Bessel function of the
first-kind and its derivatives. In this case of cylindrical
perturbed waves the fraction of energy density imparted to the
matter field is
\begin{equation}
E_{imp}=2\epsilon\frac{\psi_0'\psi_1'}{\psi_0'^2-\gamma_0'}.
\end{equation}

The non-vanishing components of the Weyl tensor for the cylindrical
wave-like spacetime are
\begin{align}
C^0_{\;\;101}&=\frac{1}{3R^5}(3R^2\rho-2R\rho^2+3\rho^3-R^3)+O(\epsilon^2),\nonumber\quad\\
C^0_{\;\;202}&=a^2C^0_{\;\;303}=-
(1+\epsilon\frac{2t}{T})\frac{a^2{e^{\rho^3/R^3}}}{6R^5}(3R^2\rho-2R\rho^2+3\rho^3-R^3)
+O(\epsilon^2),\nonumber\quad\\
C^1_{\;\;212}&=a^2C^1_{\;\;313}=-C^0_{\;\;202}, \quad
C^2_{\;\;323}=-2C^0_{\;\;202}.
\end{align}
In covariant form the components of the Weyl tensor are
\begin{align}
C_{0101}&=\frac{1}{3R^5}(1+\epsilon\frac{2t}{T})(3R^2\rho-2R\rho^2+3\rho^3-R^3)
+O(\epsilon^2),\nonumber\quad\\
C_{0202}&=a^2C_{0303}=-
\frac{a^2{e^{{\rho^3/R^3}+{\rho^2/R^2}}}}{6R^5}(1+\epsilon\frac{4t}{T})
(3R^2\rho-2R\rho^2+3\rho^3-R^3)
+O(\epsilon^2),\nonumber\quad\\
C_{1212}&=a^2C_{1313}=-C_{0202}, \quad C_{2323}=-2C_{0202}.
\end{align}
The components in the covariant form are physically reasonable as
they follow the geometry of the constructed metric and the energy
defined by approximate symmetry.

The nonzero components of stress-energy tensor are
\begin{align}
T_{00}&=\frac{3e^{\rho^2/R^2}\rho}{2\kappa R^6}(1+\epsilon
\frac{2t}{T})(4R^3+9\rho^3)+O(\epsilon^2),\quad
T_{11}=\frac{3}{2\kappa R^6}(3\rho+4R)+O(\epsilon^2),\nonumber\\
T_{22}&=a^2T_{33}= \frac{-a^2e^{\rho^3/R^3}\rho}{2\kappa
R^6}(1+\epsilon \frac{2t}{T})(4\rho^2R^2+6R^3\rho
+9\rho^4+6R\rho^3+6R^4)+O(\epsilon^2),\nonumber\\
T_{01}&=\epsilon\frac{\rho}{\kappa TR^3}(3\rho-2R)+O(\epsilon^2).
\end{align}
Here the momentum density is along radius of the cylinder. For this
case we have the same relative energy density imparted to the matter
field, as given by (42).

\section{Approximate Noether symmetries of the\\
cylindrical wave spacetimes}

The Lagrangian of the spacetime (68) is
\begin{equation}
L=e^{({\rho/R})^2}\dot{t}^2-\dot{\rho}^2-e^{({\rho/R})^3}(a^2\dot{\phi}^2+\dot{z}^2),
\end{equation}
yields the following symmetry generators
\begin{equation}
\quad \mathbf{X}_{0}=\frac{\partial}{\partial t},\quad
\mathbf{X}_{1}=\frac{\partial}{\partial \phi},\quad
\mathbf{X}_{2}=\frac{\partial}{\partial z},\quad
\mathbf{X}_{3}=z\frac{\partial}{\partial
\phi}-a^2\phi\frac{\partial}{\partial z},\quad
\mathbf{Y}_{0}=\frac{\partial}{\partial s},\quad A=c ,
\end{equation}
where $c$ is a constant, $\mathbf{X}_{0}$ corresponds to energy
conservation, $\mathbf{X}_{1}$ corresponds to azimuthal angular
momentum conservation and $\mathbf{X}_{2}$ to linear momentum
conservation along $z$, while $\mathbf{X}_{3}$ corresponds to
angular momentum conservation.

The first-order perturbed Lagrangian for the cylindrical wave-like
spacetime is
\begin{equation}
\quad
L=e^{({\rho/R})^2}\dot{t}^2-\dot{\rho}^2-e^{({\rho/R})^3}(a^2\dot{\phi}^2+\dot{z}^2)+
\frac{2\epsilon
t}{T}[e^{({\rho/R})^2}\dot{t}^2-e^{({\rho/R})^3}(a^2\dot{\phi}^2+\dot{z}^2)]
+O(\epsilon^2).
\end{equation}
For this Lagrangian along with the exact symmetry generators given
by (77), the non-trivial approximate symmetry $\mathbf{X}_a$ given
by (79) is obtained. The gauge function $A_{1}$ is again a constant,
\begin{equation}
\mathbf{X}_a= \frac{\partial}{\partial t}-\epsilon \frac{1}{T}(t
\frac{\partial} {\partial t}+\phi\frac{\partial}{\partial \phi}+z
\frac{\partial}{\partial z}).
\end{equation}
The corresponding stable first integral is
\begin{equation}
I=2e^{(\rho/R)^2}\dot{t}+\frac{2\epsilon}{T}[e^{(\rho/R)^2}t\dot{t}
+e^{(\rho/R)^3}(a^2\phi\dot{\phi}+z\dot{z})].
\end{equation}
As for the plane wave-like spacetime the following conserved
quantity corresponding to (79) is calculated
\begin{equation}
Q=E-\frac{\epsilon}{T}(t\dot{E}+\phi\dot{p_\phi}+z\dot{p_z}).
\end{equation}

Using the Lagrangian for the spacetime (54)
\begin{equation}
L=e^{2(\gamma -\psi
)}(\dot{t}^2-\dot{\rho}^2)-\rho^2e^{-2\psi}\dot{\phi}^2-e^{2\psi}\dot{z}^2,
\end{equation}
in (6) and solving the resulting system of determining equations we
obtain the symmetry generator ${\partial}/{\partial s}$ along with
the two KVs ${\partial}/{\partial \phi}$, ${\partial}/{\partial z}$
and the gauge function $A$ is a constant.

The Lagrangian for the spacetime (58) is
\begin{equation}
L=e^{2(\gamma_0 -\psi_0
)}(\dot{t}^2-\dot{\rho}^2)-\rho^2e^{-2\psi_0}\dot{\phi}^2-e^{2\psi_0}\dot{z}^2,
\end{equation}
which admits the following 4 symmetry generators along with the
gauge function as a constant
\begin{equation}
\mathbf{X}_{0}=\frac{\partial}{\partial
t},\quad\mathbf{X}_{1}=\frac{\partial}{\partial
\phi}\quad\mathbf{X}_{2}=\frac{\partial}{\partial
z},\quad\mathbf{Y}_{0}=\frac{\partial}{\partial s}.
\end{equation}

The first-order perturbed Lagrangian for the cylindrical wave
spacetime (54) with $\psi$ and $\gamma$ defined by (60) is given by
\begin{align}
L=e^{2(\gamma_0 -\psi_0
)}(\dot{t}^2-\dot{\rho}^2)-\rho^2e^{-2\psi_0}\dot{\phi}^2-e^{2\psi_0}\dot{z}^2
-2\epsilon\psi_1[e^{2(\gamma_0 -\psi_0
)}(\dot{t}^2-\dot{\rho}^2)-\rho^2e^{-2\psi_0}\dot{\phi}^2\nonumber\\
+e^{2\psi_0}\dot{z}^2]+O(\epsilon^2).
\end{align}
For $\epsilon=0$ this Lagrangian reduces to the Lagrangian (83).
Using this perturbed Lagrangian and the 4 dimensional exact symmetry
algebra of the static spacetime (58) in (16), we get a set of
determining equations in which only 1 constant corresponding to the
exact symmetry generator appears. This exact symmetry generator has
to be eliminated for consistency of these determining equations,
making them homogeneous. The resulting system is once more the same
as for the spacetime (58), yielding first-order trivial approximate
symmetry generators. Thus there is no non-trivial approximate
symmetry for this perturbed Lagrangian and the gauge function $A$ is
a constant. Hence energy conservation, azimuthal angular momentum
conservation and linear momentum conservation along the axis of the
cylinder are obtained as trivial first-order approximate
conservation laws. (Note that the technically ``trivial'' law may be
physically non-trivial.)

\section{Summary and discussion}

The problem of energy in gravitational wave spacetimes using
approximate Lie symmetry methods for DEs is addressed. To resolve
this problem we used the second-order approximate symmetries of the
geodesic equations for perturbed gravitational wave spacetimes
discussed here. First the pp-wave spacetime is investigated. Since
there is no $\epsilon^2$ in the geodesic equations for the perturbed
pp-waves, the definition of second-order approximate symmetries of
ODEs which gives the scaling factor, can not be applied to them.
This is similar to the result of Qadir and Sharif's work \cite{QS},
using the pseudo-Newtonian formalism, which just gave a constant
momentum imparted to test particles in the path of the waves and no
determinable value for it. For a better understanding of the
implication of the definition of second-order approximate symmetries
of ODEs, in plane symmetric waves this definition has applied to the
artificially constructed time-varying non-vacuum plane symmetric
spacetime \cite{IA}, for which the scaling factor (35) is obtained.
It is seen from the plots of the plane symmetric wave-like spacetime
that the energy increases with time close to the origin for $x$ and
then disappears. Then we investigated the second-order approximate
symmetries of the geodesic equations for the cylindrical wave
spacetimes. The scaling factors (67) and (69) are obtained for these
spacetimes. In the factor (67) the magnitude of the coefficient of
$(\omega\rho)^{-1/2}$ is greater then the magnitude of the
coefficient of $(\omega\rho)^{-3/2}$, therefore, the contribution of
the second term is very small and is neglected. This factor is
plotted for different values of $t$, $\rho$ and $\omega$. It shows a
behavior much more recognisably wave-like. In Figs. 3 to 6 the
energy oscillates between positive and negative values along $t$ and
$\rho$. It disappears as $\rho$ tends to infinity.

To obtain the pure gravitational field and the matter field the
approximate Weyl and stress-energy tensors for the gravitational
wave spacetimes are calculated. The components of the Weyl tensor
are given in the (0, 4) (covariant) form as well. For the perturbed
pp-wave spacetime it appears that the (0, 4) form gives the
physically relevant quantities as the space dependence in the (1, 3)
(mixed) form of the Weyl tensor does not seem to correspond to the
geometry of the pp-wave, while the covariant form does. For the
wave-like spacetimes the components in the covariant form are
physically reasonable as they follow the geometry of the constructed
metrics and the energy defined by approximate symmetry. The
stress-energy tensor density imparted to the matter field in the
wave-like and perturbed cylindrical wave spacetimes was obtained.

In GR, different people have tried to introduce the concept of a
pseudo-tensor, to define energy and momentum. In this regard first
Einstein obtained a pseudo-tensor to define energy in GR \cite{En}.
Following Einstein's idea, Landau-Lifshitz \cite{LL}, Papapetrou
\cite{PP} and Weinberg \cite{Wb} gave different pseudo-tensors to
represent the energy and momentum of the gravitational field. The
idea, of introducing a pseudo-tensor has been criticized because all
the pseudo-tensors are coordinate dependent and hence {\it
non-tensorial}. This violates the basic principles of GR. Because of
the coordinate dependence, many others, including M\"{o}ller
\cite{Mol,Mol1}, Komar \cite{Kom}, Ashtekar-Hansen \cite{AH} and
Penrose \cite{RP}, have proposed coordinate independent definitions.
M\"{o}ller realized that the use of a tetrad as the field variable,
instead of a metric, makes it possible to introduce a first order
scalar Lagrangian for the EFEs. Komar introduced a tensorial
super-potential which is independent of any background structure and
has uniqueness property. Ashtekar and Hansen defined the angular
momentum in their specific conformal model of the spatial infinity
as a certain 2-surface integral near infinity. Penrose defined
quasi-local energy-momentum and angular momentum using
twistor-theoretical idea. However, each of these, has its own
drawbacks \cite{Mol1,BG,BT}. A detailed discussion on different
definitions of gravitational energy is available in ``Quasi-local
energy-momentum and angular momentum in GR: a review article''
\cite{LBS} and ``Energy and momentum in GR'' \cite{MS}. Using the
idea of pseudo-tensors different people claimed that the
gravitational energy should be positive at large scales as well as
at small scales \cite{SY,NM,LLS1,CN1,CN2}. The positivity of
gravitational energy does not seems convincing because the total
energy of the universe is zero \cite{SYF}, which suggests that the
gravitational energy must fluctuate between positive and negative
values to give the net energy of a spacetime zero.

Our definition of gravitational energy, obtained from approximate
Lie symmetry methods, avoids the pseudo-tensor and hence does not
violate GR. Our expression of energy is also reasonable as the
gravitational energy oscillates over positive and negative values,
as it should. Admittedly, in the artificial example we constructed
the energy increased linearly without limit. This was because of the
(nonphysical) choice of a linearly increasing component of the
metric tensor for convenience of computation, leading to a
corresponding increase in the scaling factors (35) and (69). For the
physical example of cylindrical exact waves, the Bessel function of
the first kind goes to zero asymptotically for large values of the
argument \cite{NRS}. Correspondingly, our scaling factor (67) for
cylindrical waves dies out asymptotically, giving a net zero energy.

One of us (AQ) was drawn to study the energy of gravitational waves
when a question was posed \cite{SMM} whether there is the analogue
of Landau-damping of electromagnetic waves for gravitational waves.
Since Maxwell's theory of electromagnetism is linear,
electromagnetic waves do not interact with the field but are damped
due to their interaction with matter. On the other hand GR is
non-linear and so gravitational waves undergo self-interaction. This
gives rise to the possibility of ``Landau self-damping" of
gravitational waves. On the other hand, the Khan-Penrose \cite{KP}
and Szekeres \cite{PS} solutions of colliding plane gravitational
waves suggest that there could even be {\it enhancement} of the
waves, as they lead to curvature singularities after the collision.
The problem of definition of energy in GR makes it very difficult to
answer the question posed. Using Wheeler's ``poor man's approach",
we can ask whether ``the mass equivalent to the energy of the
gravitational wave attracts and hence damps the waves", or like the
black hole, ``the energy enhances the mass and hence the energy
equivalent to it in the wave". With our present proposal the
question seems to be answerable. Classically the energy density in
cylindrical waves reduces by the factor $1/(2\pi\rho)$. From (67)
the energy density decreases by a further factor of
$(3\times2^{11/4})/\sqrt{\pi^3\times(\omega\rho)}$. Hence for
sufficiently large $\rho$ the scaling factor $\sim
1/\sqrt{\omega\rho^{3}}$ which is a significant {\it self-damping}
of the waves! This enhanced asymptotic attenuation of gravitational
waves will obviously have profound observational significance.

It would be of great interest to apply this approximate symmetry
analysis to the Khan-Penrose and Szekeres solutions to see whether
they suffer self-damping or enhancement according to our definition.
Of course, it may be that the procedure will be inapplicable for
those plane wave solutions as well. Also, the analysis should be
applied to ``spherical solutions" like those of Nutku \cite{YN}.

\section*{Acknowledgments} We thank F. De Paolis for pointing out a
serious error in an earlier draft of this paper. IH would like to
thank the Higher Education Commission of Pakistan (HEC) for their
full financial support and DECMA, where this work was partially
done.


\end{document}